\documentclass[journal,draftclsnofoot,onecolumn,12pt]{IEEEtran}
\usepackage[utf8]{inputenc}
\usepackage{amsmath}
\usepackage{amssymb}
\usepackage{graphicx}
\usepackage{multirow}
\usepackage{cite}
\usepackage{xcolor}

\usepackage[capitalise,noabbrev]{cleveref}
\Crefformat{figure}{#2Fig.~#1#3}
\Crefformat{equation}{#2(#1)#3}

\usepackage[detect-all,range-phrase=--,per-mode=symbol,range-units=single,list-units=single]{siunitx}
\DeclareSIUnit\dBm{dBm}
\DeclareSIUnit\dB{dB}
\DeclareSIUnit\dBi{dBi}
\title{Impact of Passive Element Technological Limits on CMOS Low-Noise Amplifier Design
\thanks{This document is the author's translation of a peer-reviewed paper published initially in Spanish. \textbf{How to cite}: J. L. Gonz\'alez, R. L. Moreno, and D. V\'azquez, "L\'imites impuestos por los elementos pasivos en el diseño de amplificadores de bajo ruido en tecnolog\'ia CMOS," Revista de Ingenier\'ia Electr\'onica, Autom\'atica y Comunicaciones, vol. 36, no. 3, pp. 1-12, 2015. [Online]. Available: http://rielac.cujae.edu.cu/index.php/rieac/article/view/296.}
}

\author{
    \IEEEauthorblockN{J. L. Gonz\'alez, R. L. Moreno, D. V\'azquez}
}

\begin{document}

\maketitle

\begin{abstract}
This paper investigates the impact of technological constraints on passive elements in the design of inductively degenerated CMOS low-noise amplifiers (LNAs). A theoretical analysis is combined with circuit simulations in a 130-nm CMOS process at 2.45~GHz to explore how the available inductance and capacitance values limit key design objectives such as maximum gain, minimum power consumption, and transistor sizing. Results show that these limits significantly restrict the achievable design space, particularly for low-power implementations, and highlight the need to incorporate detailed passive-element models into RF integrated circuit design flows.
\end{abstract}


\section{Introduction}

In 1958, Jack Kilby demonstrated that both transistors and passive elements could be fabricated on a single semiconductor substrate, giving rise to the integrated circuit~\cite{millman1993}. This milestone enabled rapid advances in microelectronics, driving the development of communication, information, and computing technologies that are essential in modern society.

The integrated circuit market is largely dominated by CMOS technology due to its low production cost~\cite{baker2010}. Continuous MOS transistor scaling has increased processing capability and speed while reducing power consumption~\cite{sah1988}. These improvements benefit both digital and analog circuits, including those for radio-frequency (RF) applications~\cite{woerlee2001}. Current submicron CMOS technologies allow the integration of complete communication systems—from RF front-ends to digital baseband processing—on a single chip, enabling wireless communication at gigahertz (GHz) frequencies~\cite{vidojkovic2008, schneider2010}. However, despite their presence since the inception of integrated circuits, the fabrication and modeling complexity of passive elements—particularly inductors—remains a major challenge in RF design~\cite{lee2000}.

In integrated communication systems, RF block design is a critical challenge~\cite{razavi1998}. For RF receivers, the performance of the low-noise amplifier (LNA) strongly influences overall system behavior~\cite{razavi1998, lee2004}. The LNA must ensure sufficient gain and low noise contribution to achieve the required receiver sensitivity~\cite{leroux2005}. Additionally, it must provide proper input impedance matching (typically to a 50~$\Omega$ source), high linearity—commonly characterized by the third-order intermodulation intercept point (IP3)—and good reverse isolation~\cite{lee2004}.

\begin{figure}[!htbp]
\centering
\includegraphics[width=0.7\linewidth]{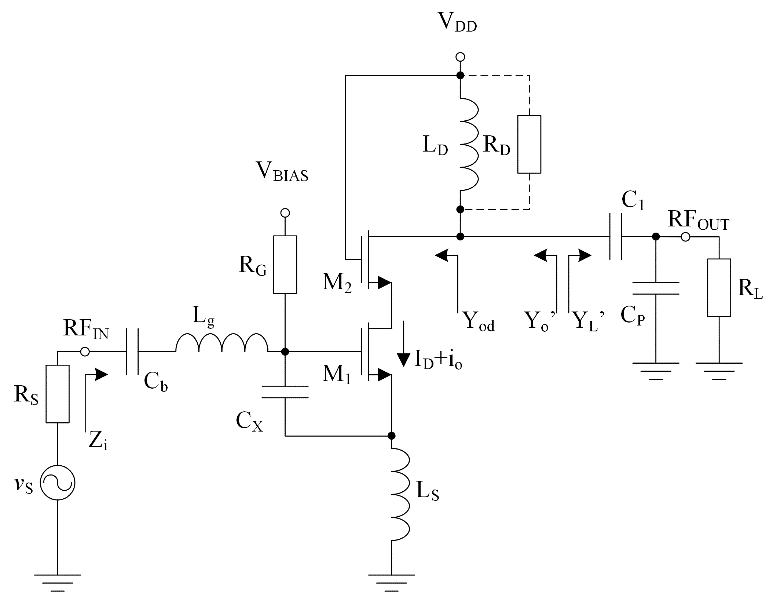}
\caption{Common-source CMOS LNA with inductive degeneration.}
\label{fig:lna}
\end{figure}

The inductively degenerated common-source LNA (CS-LNA), shown in Fig.~\ref{fig:lna}, is widely used in CMOS receivers for short-range wireless standards such as Wi-Fi, Bluetooth, and ZigBee~\cite{lee2004, leroux2005, farahani2008}. For a given gain and power budget, transistor sizing can minimize the noise figure (NF)~\cite{shaeffer1997,andreani2001,belostotski2006}. Furthermore, high IP3 with low power can be achieved by exploiting the linearity sweet spot that occurs when MOS transistors operate in moderate inversion~\cite{aparin2004,niu2005,toole2004}. This peak is associated with a specific current density in the common-source device, making transistor sizing critical for optimizing the trade-off among noise, linearity, and power consumption.

To achieve this balance, a design space exploration is proposed, consisting of sweeping the width of $M_1$ ($W_1$) for different bias currents ($I_D$) \cite{gonzalez2016ieeelat,gonzalez2016apec,gonzalez2016informatica}. For each combination, the passive element dimensions are synthesized to meet gain and impedance matching requirements, after which NF and IP3 are evaluated. The resulting set of designs defines the design space from which an implementation can be selected. However, the synthesis of each LNA is constrained by the technological limits of passive elements, which impose additional restrictions on achievable performance—a topic rarely addressed in the literature.

This work analyzes these constraints for a CS-LNA implemented in a 130-nm CMOS process with a 1.2-V supply and a target frequency of 2.45~GHz. First, qualitative dependencies between passive element values and design objectives (gain, bias current, transistor width) are derived from circuit analysis. These are then validated and refined through simulations. Results show that maximum gain, minimum bias current (and thus minimum power), and transistor dimensions are ultimately limited by the inductors and capacitors available in the technology.

\section{Theoretical Analysis of Passive Element Constraints}

\subsection{Description and Analysis of the Common-Source LNA with Inductive Degeneration}

Figure~\ref{fig:lna} shows the basic schematic of a common-source (CS) LNA with inductive degeneration. The source inductor $L_S$ introduces a resistive component in the input impedance without adding extra noise sources \cite{karanicolas1996}. The capacitor $C_X$ allows minimizing the noise figure for specific values of gain and power consumption \cite{shaeffer1997}. The gate inductor $L_g$ is included to tune the input impedance. Transistor M2 is used as a cascode stage to reduce the Miller effect on M1 and to improve reverse isolation \cite{lee2004}. At the output, $L_D$ forms a parallel resonant network with the output capacitances of the cascode stage and the impedance seen toward the load. A capacitive divider $(C_1, C_P)$ is included to couple the output impedance to $50~\Omega$ for stand-alone LNA characterization with a spectrum analyzer. The gate-bias resistor ($R_G$) and the input DC-blocking capacitor ($C_b$) must present sufficiently high and low impedances, respectively, such that their effects are negligible under normal operating conditions. The gate bias voltage of M1, $V_{\mathrm{BIAS}}$, can be derived from $V_{DD}$ by means of a current mirror or another voltage reference circuit \cite{razavi1998}.

The available power gain of the LNA, $G$, can be written as a function of the input-stage transconductance $G_m \!\triangleq\! \lvert I_o/V_s\rvert$\footnote{$V_s$ and $I_o$ denote RMS values.}, the source resistance $R_S$, and the output-stage conductance $G_o' \!=\! \operatorname{Re}\{Y_o'\}$. Assuming impedance matching at both ports and neglecting losses in the output coupling capacitors, one obtains
\begin{equation}
G \triangleq \frac{P_o}{P_{\text{avs}}}
= \frac{I_o^2/(4\,G_o')}{V_s^2/(4\,R_S)}
= G_m^2\frac{R_S}{G_o'} .
\label{eq:G_def}
\end{equation}

The output-stage conductance is the parallel combination of the cascode output conductance and the inductor's parallel loss. Denoting by $Y_{od}$ the small-signal output admittance of the cascode stage and by $R_D$ the equivalent parallel resistance of $L_D$, we have
\begin{equation}
G_o' \equiv \operatorname{Re}\{Y_o'\}
= \operatorname{Re}\{Y_{od}\} + \frac{1}{R_D},
\qquad
R_D = \omega_0 L_D Q_D ,
\label{eq:Go_def}
\end{equation}
where $\omega_0$ is the operating angular frequency and $Q_D$ is the quality factor of $L_D$.

Using a simplified small-signal analysis of the input stage (considering in M1 only the gate-source capacitance $C_{gs}$ and the controlled current source $i_o = g_m v_{gs}$, and treating $L_S$, $L_g$, and $C_X$ as ideal while neglecting the loading of the cascode on the input), the effective transconductance under input matching can be approximated as \cite{lee2000}
\begin{equation}
G_m \simeq \frac{1}{2\,\omega_0\,L_S}.
\label{eq:Gm_simple}
\end{equation}

Equations~\eqref{eq:G_def}--\eqref{eq:Gm_simple} show that the LNA gain depends on both the drain inductor $L_D$ and the source-degeneration inductor $L_S$: on the output side through the product $L_D Q_D$ (via $R_D$), and on the input side mainly through its inductance. When selecting $L_D$, the frequency response of the output impedance must also be taken into account. To widen the frequency range over which the output impedance remains adequate, the quality factor of the resonant network should be reduced; thus, a lower $Q_D$ must be chosen. To maintain the same contribution of the output network to the gain when decreasing $Q_D$, $L_D$ must be increased proportionally. However, the maximum achievable inductance is limited both by the technological constraints of the on-chip inductors (including the dependence of feasible $Q$ values on inductance \cite{andreani2001}) and by the minimum capacitance in the output resonant network \cite{belostotski2006}. Due to these trade-offs and constraints, it is convenient to fix the characteristics of $L_D$ first and then select the remaining passive elements\cite{gonzalez2016informatica}.

\begin{figure}[!htbp]
    \centering
    \includegraphics[width=0.6\linewidth]{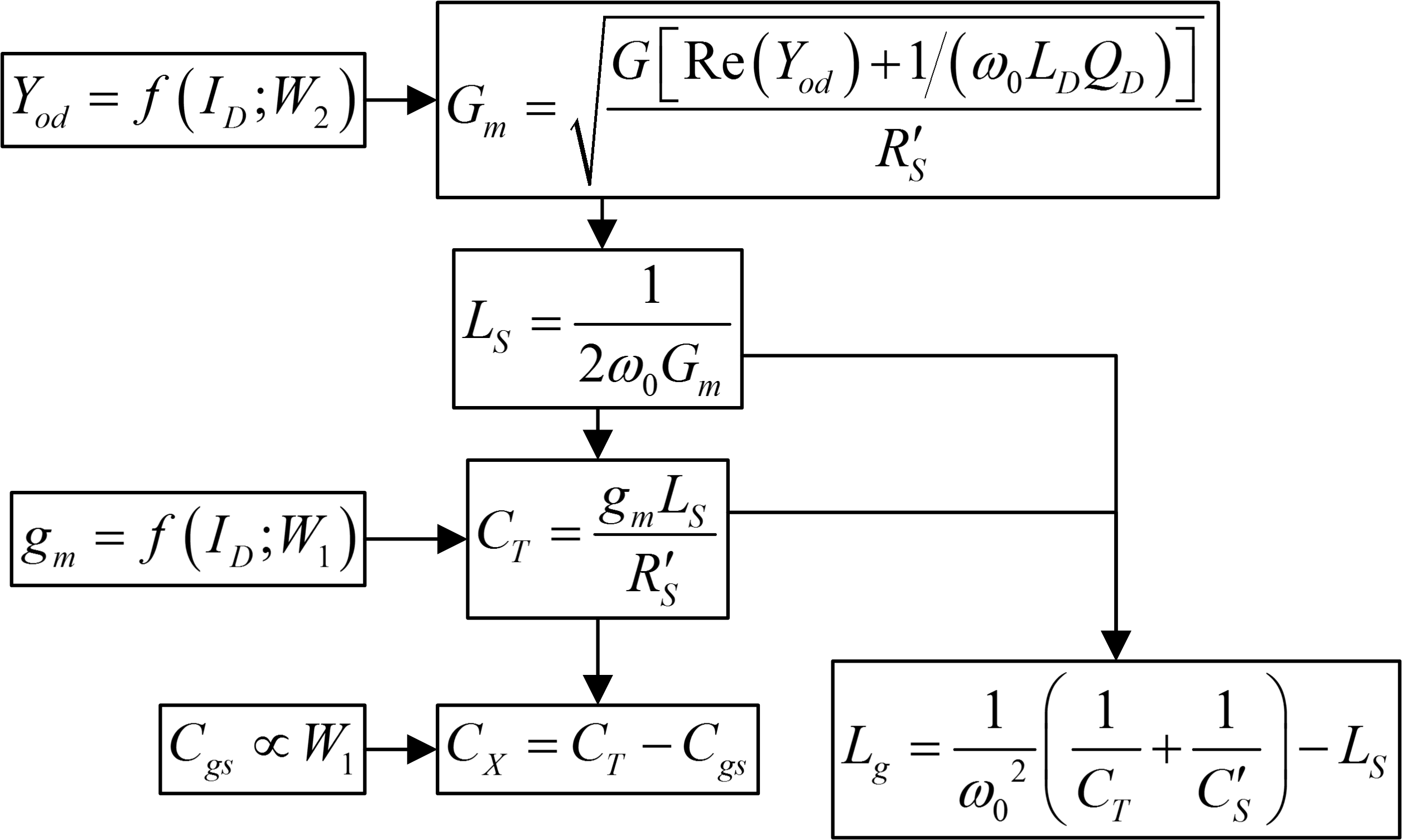}
    \caption{Dependence of the passive elements in the transconductance stage on the LNA gain and the transistor dimensions and biasing, according to the analysis using simplified models.}
    \label{fig:flow_diagram}
\end{figure}

The previous set of equations obtained from the simplified small-signal analysis establishes a logical sequence for determining the passive elements of the input stage. This sequence is depicted as a flowchart in Fig.~\ref{fig:flow_diagram}, which includes the dependencies on device sizing and biasing \cite{gonzalez2016informatica}. Here, $C_T = C_X + C_{gs}$ denotes the total equivalent capacitance between the gate and the source of M1. Due to the modeling simplifications, these expressions are not accurate enough to compute final design values; however, they are very useful for anticipating the restrictions imposed by technological limits.

\subsection{Behavior of Passive Elements as a Function of Synthesis Objectives}

\subsubsection{Increasing the Contribution of the Transconductance Stage to the Gain (While Keeping Bias Current and Transistor Width Constant)}

The fundamental contribution to the LNA gain must be provided by the input stage, through a sufficiently large value of $G_m$, in order to reduce the noise figure \cite{lee2004,leroux2005}. According to the expressions shown in Fig.~\ref{fig:flow_diagram}, to increase $G_m$ one must reduce the value of $L_S$, which in turn implies a reduction of $C_T$ to keep the input resistance constant (taking into account that the transistor transconductance, $g_m$, does not change for constant $W_1$ and $I_D$). Both the reduction of $L_S$ and of $C_T$ require an increase of $L_g$ in order not to alter the resonance frequency. Moreover, for fixed dimensions of M1 (constant $C_{gs}$) the reduction of $C_T$ is obtained by using a smaller value of $C_X$.

Therefore, the maximum value of $G_m$ can be limited by the minimum feasible value of $L_S$ ($L_{S,\text{min}}$), the minimum available capacitance for $C_X$ ($C_{X,\text{min}}$), or the maximum achievable value of $L_g$ ($L_{g,\text{max}}$).

\subsubsection{Decreasing the Bias Current (While Keeping Gain and Transistor Width Constant)}

When the bias current $I_D$ is reduced, the transconductance of the transistor $g_m$ also decreases. In order to maintain the same effective transconductance $G_m$ (and therefore preserve the gain), the degeneration inductance $L_S$ must be increased. According to the relationships summarized in Fig.~\ref{fig:flow_diagram}, an increase in $L_S$ requires a simultaneous increase in $C_T$ in order to keep the input resistance constant. Since the width of the transistor remains fixed (hence $C_{gs}$ is constant), this increase in $C_T$ must be achieved by using a larger value of $C_X$. At the same time, in order not to alter the resonance frequency, $L_g$ must be reduced.

Therefore, the minimum feasible bias current is limited by the maximum values that can be achieved for $L_S$ and $C_X$, and by the minimum achievable value of $L_g$.

\subsubsection{Limits on the Transistor Width (While Keeping Gain and Bias Current Constant)}

In the proposed LNA design methodology, the transistor width is swept while keeping the gain and the bias current fixed. To analyze the impact of width variations, we again assume a constant transconductance $G_m$ (and therefore a constant $L_S$). Reducing the width of $M_1$ ($W_1$) at constant current decreases its transconductance $g_m$~\cite{razavi2001}, which in turn requires reducing $C_T$. As previously discussed, this leads to an increase in $L_g$. Hence, the minimum $W_1$ may be limited by $L_{g\text{max}}$. Conversely, increasing the transistor width reduces $L_g$; however, this is unlikely to be a practical constraint, as it would imply very wide devices that are uncommon in low-power designs.

The dependence of $C_X$ on transistor width is more complex. As the width of $M_1$ increases, not only does $C_T$ increase (through its relation to $g_m$), but the intrinsic gate–source capacitance $C_{gs}$ of $M_1$ also increases. Therefore, the behavior of $C_X$ depends on the relative rate of change of $C_T$ and $C_{gs}$ with respect to transistor width: if $C_T$ increases more than $C_{gs}$ for the same increment in $W_1$, then $C_X$ increases, and vice versa. Equation \eqref{eq:cx} captures this dual dependence, under the approximations that transconductance is proportional to the square root of the product $I_D W_1$ and that gate–source capacitance is proportional to transistor width~\cite{razavi2001}. Assuming unit proportionality constants as a hypothetical case, Fig.~3 illustrates the dependence of $C_X$ on transistor width for different bias current values.

\begin{equation}
    \label{eq:cx}
    C_X = C_T - C_{gs} = \frac{g_m L_S}{R_S} - C_{gs} = k_{g_m} \sqrt{I_D W_1} - k_{C_{gs}} W_1
\end{equation}

\begin{figure}[!htbp]
    \centering
    \includegraphics[width=1\linewidth]{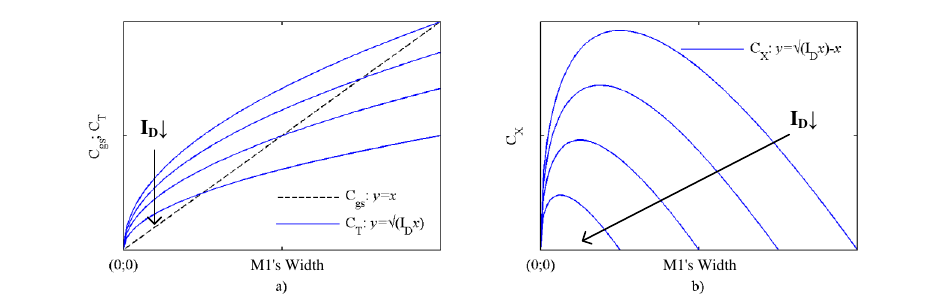}
    \caption{Simplified theoretical dependence of $C_X$ on the width of $M_1$ ($W_1$) for different values of the bias current: a) functions describing the behavior of $C_T$ [$y = \sqrt{ID \cdot x}$] and $C_{gs}$ ($y = x$); b) function describing the behavior of $C_X$ [$y = \sqrt{ID \cdot x} - x$]. The $y$-axis represents the capacitance values, while the $x$-axis represents the transistor width.}
    \label{fig:C_vs_W}
\end{figure}

As shown in Fig.~\ref{fig:C_vs_W}b, the maximum transistor width is limited by the minimum value of $C_X$. Furthermore, the lower the bias current, the smaller the maximum $W_1$, which means that reducing power consumption restricts the transistor dimensions available for LNA implementation. From the behavior illustrated in Fig.~\ref{fig:C_vs_W}b, it can also be inferred that there exists an interval of $W_1$, around the maximum of the function, where the required $C_X$ exceeds the maximum capacitance supported by the technology. However, since this occurs only at higher $I_D$ values, it should not represent a practical limitation when designing for low-power operation. Therefore, the maximum $W_1$ may ultimately be constrained by $C_{X\text{min}}$.

\subsubsection{Second-Order Effects Not Considered in the Simplified Models}

Previous studies on this technology have shown that capacitive effects of transistor $M_1$ not previously considered -- specifically, the gate-to-substrate capacitance ($C_{gb}$) and the gate-to-drain capacitance ($C_{gd}$) -— cause a reduction in power gain compared to that predicted by simplified models~\cite{leroux2005}. Similarly, reducing the bias current $I_D$ increases the input impedance of the cascode stage (the inverse of the transconductance of $M_2$, $g_{m2}$~\cite{razavi2001}), which also decreases the gain.

Therefore, the wider the transistors (resulting in larger parasitic capacitances) or the lower the bias current, the higher the required $G_m$ to compensate for the gain reduction introduced by both conditions. This implies that the minimum bias current and the maximum transistor width are also constrained by the same factors that limit the maximum achievable $G_m$, namely $L_{S\text{min}}$, $C_{X\text{min}}$, and $L_{g\text{max}}$. However, in the particular case of $L_g$, its increase due to the second-order effects of wider transistors must be counterbalanced by the decrease in inductance predicted by simplified-model analysis under the same condition.

Table~\ref{tab:limits} summarizes the influence of passive-element technological limits on LNA design, based on the theoretical analysis in this section. It includes both the primary effects predicted by simplified models and the secondary considerations discussed above.

\begin{table}[h]
\centering
\caption{Technological limits of passive elements and their influence on LNA design. Primary effects (P) and second-order effects (S).}
\label{tab:limits}
\begin{tabular}{lcccc}
\hline
 & Max $G_m$ & Min $I_D$ & Min $W_1$ & Max $W_1$ \\
\hline
Min $L_S$ & P & S & -- & S \\
Max $L_g$ & P & P+S & P & S \\
Min $C_X$ & P & P+S & -- & P+S \\
\hline
\end{tabular}
\end{table}

\section{Verification Through Simulations for a Specific Technology and Application in Design}

To validate and complement the theoretical analysis presented in the previous section, a design space exploration was performed using device models provided by the technology vendor for a 130~\text{nm} CMOS process (1P8M: one polysilicon layer and eight metal layers) with a nominal channel length of 130~\text{nm} and a supply voltage of 1.2~V.

For the LNA synthesis, the design specifications included a minimum gain of 10~\text{dB} and input/output impedance matching better than $-10$~\text{dB}, referenced to 50~$\Omega$, over the 2.4–2.5~GHz band. These requirements correspond to the implementation of a ZigBee/IEEE 802.15.4 receiver~\cite{trung-kien2006, fiorelli2011}.

The width of transistor $M_2$ ($W_2$) was set to $W_2 = W_1 / 2$ to reduce its contribution to the load capacitance, thereby improving the selection margin of the output matching network~\cite{leroux2005}. All transistors were assigned the same channel length ($L_1 = L_2 = L$), and two values were evaluated: the minimum allowed by the technology ($L_{\text{min}} = 120$~\text{nm}) and twice that value ($2L_{\text{min}} = 240$~\text{nm}).

\subsection{Characteristics of Available Passive Elements}

\begin{figure}[!htbp]
    \centering
    \includegraphics[width=1\linewidth]{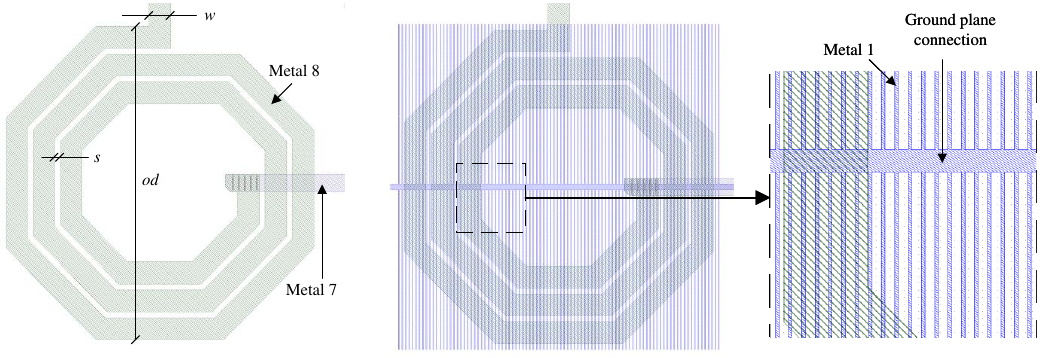}
    \caption{Structural details of an inductor from the available technology (number of turns $n_t = 2\frac{3}{4}$): (left) structural parameters of the spiral traces, (center) inductor with ground plane to mitigate substrate coupling, and (right) close-up of the ground plane.}
    \label{fig:inductor_geometry}
\end{figure}

The capacitors in the selected technology are of the metal–insulator–metal (MiM) type, while the inductors are implemented using octagonal spirals. These inductors can optionally include a ground shield to reduce coupling with the substrate~\cite{yue1998}, as illustrated in \Cref{fig:inductor_geometry}.

The parameters for circuit modeling of these passive elements were obtained through simulations based on their constructive features: for capacitors, the plate area of the parallel-plate structure (the technology fixes the separation between plates); for inductors, the number of turns ($n_t$), the outer diameter ($o_d$), and the trace width ($w$), with the spacing between turns ($s$) also fixed by the technology.

\begin{figure}[!htbp]
    \centering
    \includegraphics[width=1\linewidth]{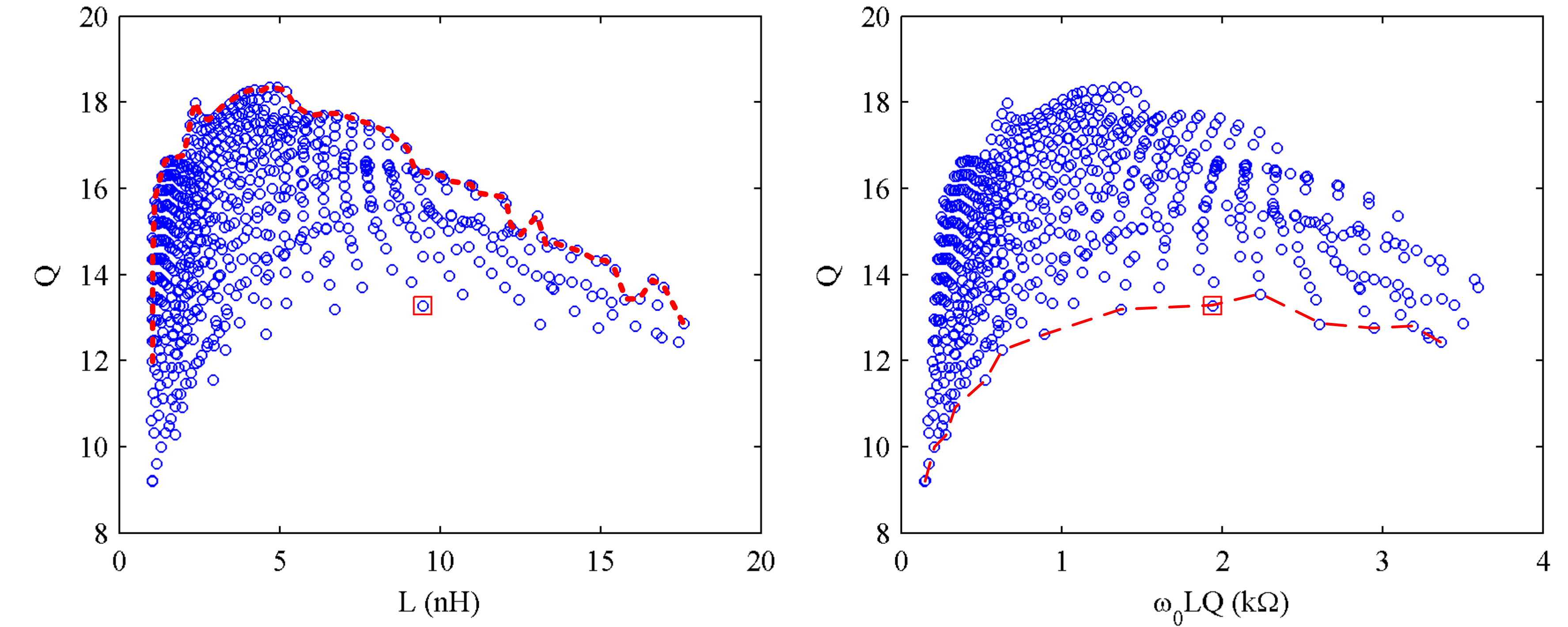}
    \caption{Characteristics of the inductors available in the technology: (left) Quality factor ($Q$) vs. inductance ($L$), where the dashed line marks the inductors with the highest $Q$. (right) Quality factor ($Q$) vs. parallel parasitic resistance ($\omega_0 L Q$), where the dashed line marks the inductors with the lowest $Q$. In both graphs, the drain inductor $L_D$ is indicated (square-enclosed circle).}
    \label{fig:inductor_design_space}
\end{figure}

Capacitances up to 5~\text{pF} can be achieved with negligible losses relative to the reactive component. For inductors, the relationships between inductance, quality factor ($Q$), and parallel parasitic resistance were analyzed and are plotted in \Cref{fig:inductor_design_space}. 

Gate and source inductors ($L_g$ and $L_S$) were selected from those with the highest quality factor in each inductance range, to minimize their contribution to the overall noise figure~\cite{lee2004, leroux2005, fiorelli2014}, as shown in \Cref{fig:inductor_design_space}a. 
The drain inductor ($L_D$) was chosen among those with lower quality factors, based on its parallel parasitic resistance (see \Cref{fig:inductor_design_space}b), specifically one with intermediate values of resistance and inductance ($L_D = 9.5$~\text{nH}, $Q_D = 13$, $R_D \approx \SI{2}{\kilo\ohm}$
).

These values allow, if required, either enhancing the gain contribution of this inductor while keeping an approximately constant bandwidth (since increasing resistance above ${\SI{2}{\kilo\ohm}}$ results in minimal variations in $Q$), or broadening the bandwidth at the expense of gain. 

\subsection{Behavior of Passive Elements According to Synthesis Objectives in the Available Technology}

The behavior of passive elements as a function of synthesis objectives was analyzed in two phases. In the first phase, a sweep of the bias current ($I_D$) and the transistor width of transistor $M_1$ ($W_1$) was performed, targeting a gain near the minimum specification ($10.5 \pm 0.5$~\text{dB}). In the second phase, $I_D$ was fixed, and a sweep of gain and $W_1$ was conducted. For each combination of gain, current, and transistor width, simulation-based optimizations determined the passive element dimensions required to meet impedance matching specifications (with an additional 5~\text{dB} margin) and the desired gain.

\subsubsection{Variation of Bias Current and Transistor Width}
\begin{figure}[!htbp]
\centering
\includegraphics[width=0.49\linewidth]{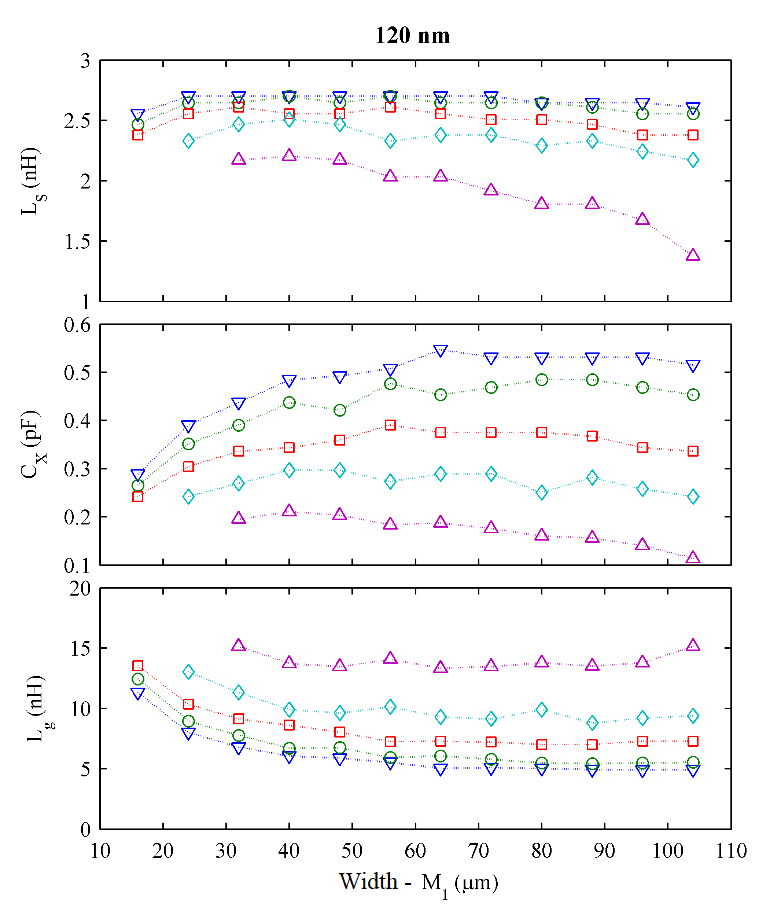}
\includegraphics[width=0.49\linewidth]{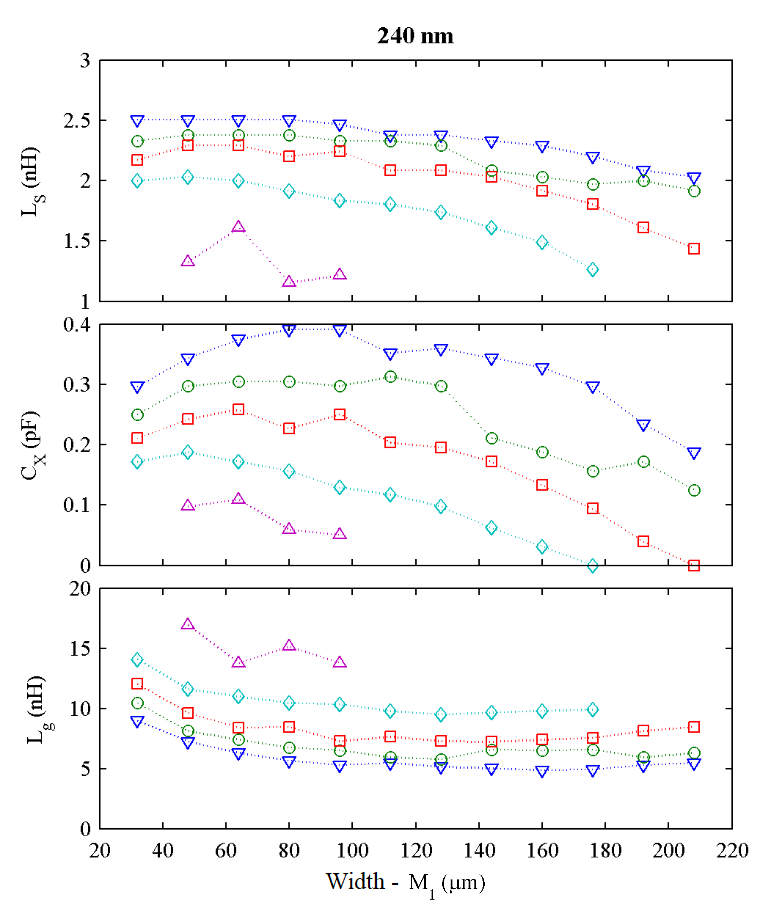}
\caption{Passive element values versus $W_1$ and $I_D$ ($\triangle$: $I_D = 0.3$~\text{mA}; $\lozenge$: $I_D = 0.4$~\text{mA}; $\square$: $I_D = 0.5$~\text{mA}; $\circ$: $I_D = 0.6$~\text{mA}; $\triangledown$: $I_D = 0.7$~\text{mA}). Target gain: $10.5 \pm 0.5$~\text{dB}.}
\label{fig:Fig6}
\end{figure}

Fig.~\ref{fig:Fig6} shows the passive element values for LNAs synthesized with a target gain of $10.5 \pm 0.5$~\text{dB} as a function of $W_1$ and $I_D$. All designs achieved input/output matching ($S_{11}, S_{22} < -15$~\text{dB}) and gains within $[10.3, 10.8]$~\text{dB}.

The results confirm both primary and secondary effects predicted by theoretical analysis. As $I_D$ decreases, the source degeneration inductance ($L_S$) must be reduced, the required gate inductor ($L_g$) increases, and $C_X$ decreases. For higher $I_D$, $L_S$ remains nearly constant with respect to $W_1$. However, at lower $I_D$, $L_S$ decreases more significantly as $W_1$ increases, especially for longer channel lengths. An additional phenomenon not predicted theoretically was observed: $L_S$ decreases for small $W_1$. This may be due to increased losses in the gate inductor (for small $W_1$, $L_g$ is larger and its quality factor decreases, increasing series resistance), which must be compensated by higher input-stage transconductance. Consequently, $L_g$ increases and $C_X$ decreases beyond theoretical predictions. For $I_D = 0.3$~\text{mA}, $L_g$ also increases with $W_1$.

The behavior of $C_X$ with respect to $W_1$ matches theoretical predictions and depends on $I_D$: $C_X$ initially increases with $W_1$, reaches a maximum, and then decreases. Both the maximum $C_X$ and the corresponding $W_1$ decrease as $I_D$ decreases. For $L = 240$~\text{nm}, $C_X$ values are lower than for $L = 120$~\text{nm} because the intrinsic gate-source capacitance ($C_{gs}$) is higher, requiring a smaller $C_X$ to maintain the total equivalent capacitance ($C_T$) and preserve input resistance. Additionally, $L_S$ exhibits greater variation with decreasing $I_D$ in longer-channel devices.

Within the analyzed ranges ($16$–$104~\mu\text{m}$ for $L = 120$~\text{nm} and $32$–$208~\mu\text{m}$ for $L = 240$~\text{nm}), technological limits constrain LNA synthesis. For both channel lengths, at $I_D = 0.3$~\text{mA}, the required $L_g$ approaches the maximum available value (18~\text{nH}), preventing designs with lower power consumption. Furthermore, the maximum $L_g$ limits the minimum $W_1$ for low $I_D$. For $L = 240$~\text{nm}, the minimum limits of $L_S$ (1~\text{nH}) and $C_X$ (0~\text{pF}) are also reached as $W_1$ increases at low $I_D$, reducing the number of feasible designs.

\subsubsection{Variation of LNA Gain and Transistor Width}

\begin{figure}[!htbp]
\centering
\includegraphics[width=0.49\linewidth]{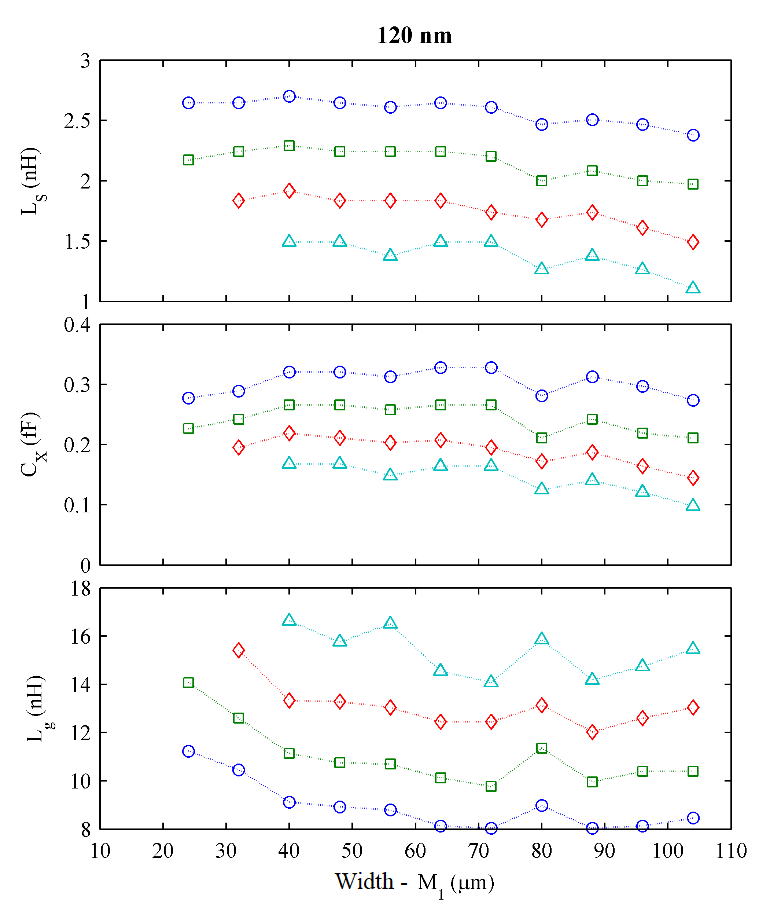}
\includegraphics[width=0.49\linewidth]{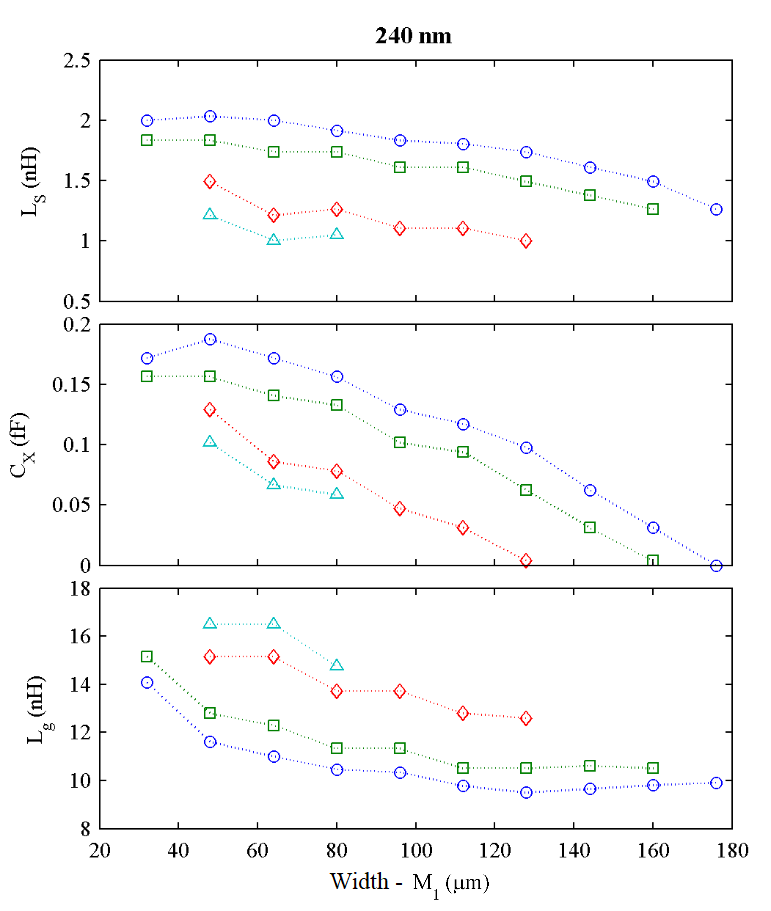}
\caption{Passive element values versus $W_1$ and LNA gain ($\circ$: $G = 10.5$~\text{dB}; $\square$: $G = 11$~\text{dB}; $\lozenge$: $G = 12$~\text{dB}; $\triangle$: $G = 13$~\text{dB}). Bias current: 0.4~\text{mA}.}
\label{fig:Fig7}
\end{figure}

To analyze the dependence of passive elements on LNA gain, the bias current was fixed at 0.4~\text{mA}, and amplifiers were synthesized for different target gains, starting from 10.5~\text{dB}. Fig.~\ref{fig:Fig7} shows the passive element values for each synthesized LNA as a function of $W_1$ and the desired gain. All designs achieved input/output matching ($S_{11}, S_{22} < -15$~\text{dB}) and gains within $\pm 0.3$~\text{dB} of the nominal value.

The results confirm the dependencies predicted by circuit analysis: increasing gain requires reducing $L_S$ and $C_X$, while increasing $L_g$. For both channel lengths, the maximum achievable gain is limited by the smallest physically realizable degeneration inductor. As gain increases, the number of feasible designs decreases due to constraints imposed by the maximum $L_g$ on the minimum transistor width and by the minimum $L_S$ and $C_X$ on the maximum width. This reduction is more pronounced for longer channel lengths.

The variation of passive elements with respect to $W_1$ for each gain value is consistent with theoretical predictions and the trends observed in the previous subsection.

\subsection{Design Space Exploration}

\begin{figure}[!htbp]
\centering
\includegraphics[width=0.49\linewidth]{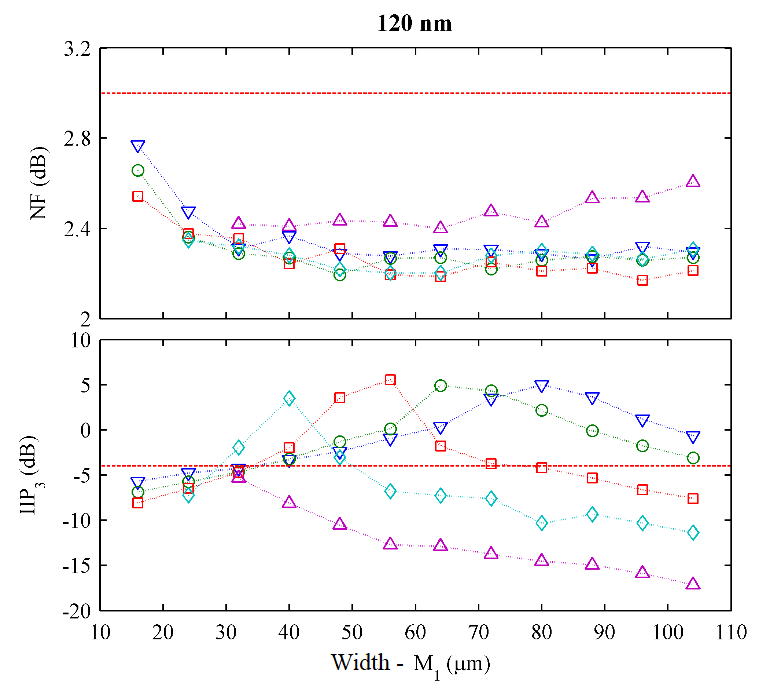}
\includegraphics[width=0.49\linewidth]{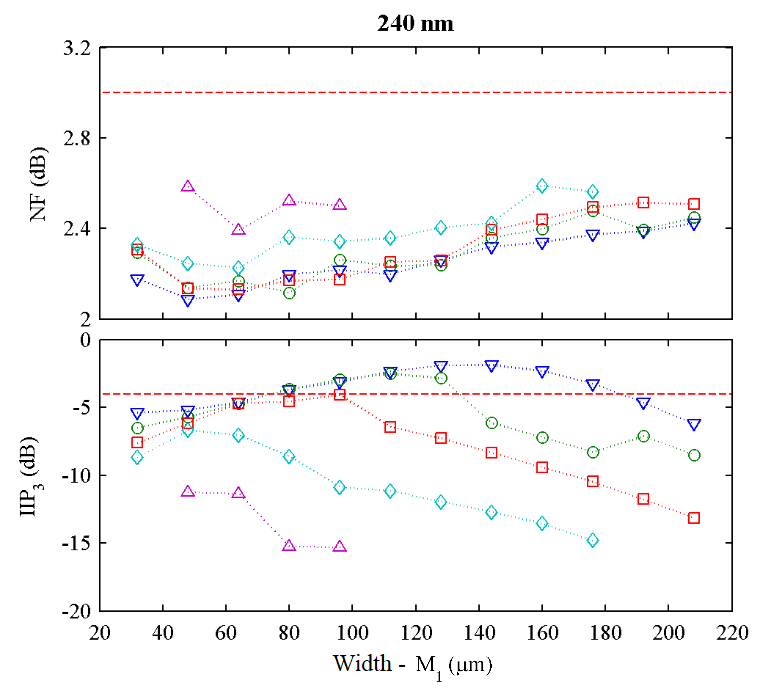}
\caption{Simulation results at 2.45~GHz for NF (top) and IIP3 (bottom) versus $W_1$ and $I_D$ ($\triangle$: $I_D = 0.3$~\text{mA}; $\lozenge$: $I_D = 0.4$~\text{mA}; $\square$: $I_D = 0.5$~\text{mA}; $\circ$: $I_D = 0.6$~\text{mA}; $\triangledown$: $I_D = 0.7$~\text{mA}). Target gain: $10.5 \pm 0.5$~\text{dB}. Horizontal dashed lines indicate ZigBee requirements.}
\label{fig:Fig8}
\end{figure}

Fig.~\ref{fig:Fig8} shows the simulation results at 2.45~GHz for the noise figure (NF) and the input-referred third-order intercept point (IIP3) of LNAs synthesized with a target gain of $10.5 \pm 0.5$~\text{dB}, completing the design space exploration. The horizontal dashed lines indicate the requirements for a ZigBee receiver~\cite{trung-kien2006, fiorelli2011}. All synthesized LNAs meet the NF specification ($\text{NF} < 3$~\text{dB}), but the required linearity ($\text{IIP3} > -4$~\text{dB}m) is not satisfied for the lowest bias currents ($I_D = 0.3$~\text{mA} with 120~\text{nm} transistors and $I_D < 0.5$~\text{mA} with 240~\text{nm} transistors).

For $I_D = 0.3$~\text{mA}, neither transistor type reaches the IIP3 peak, as only the decreasing region of this parameter is observed with increasing $W_1$. This occurs because circuits with narrower transistors could not be synthesized due to the limitations imposed by the available passive elements. For LNAs with 120~\text{nm} transistors, the IIP3 peaks for $I_D \geq 0.4$~\text{mA} significantly exceed the required linearity, suggesting that similar performance could be expected for $I_D = 0.3$~\text{mA} if designs with $W_1 < 32~\mu$\text{m} were feasible.

Therefore, the technological limits of passive elements directly and indirectly constrain the minimum power consumption that can be achieved in the design of this type of amplifier.

\section{Conclusions}


This work analyzed the constraints imposed by the technological limits of passive elements on the design of inductively degenerated CMOS LNAs. The study demonstrated that maximum gain, minimum bias current (and thus minimum power), and transistor dimensions are strongly determined by the extreme values of inductance and capacitance available in the technology. Reducing the channel length increases the number of feasible designs, but does not eliminate these limitations.

These findings underscore the importance of integrating accurate passive-element models and their technological limits into RF design methodologies. Future work will focus on developing automated design tools that account for these constraints and on exploring alternative topologies or passive-element implementations to further expand the achievable design space.

\section*{Acknowledgments}
This work was supported by CAPES-Brazil (Project 176/12), CNPq, MAEC-AECID (Project FORTIN, Ref. D/024124/09), FEDER Program of Junta de Andalucía (Project P09-TIC-5386), and the Spanish Ministry of Economy and Competitiveness (Project TEC2011-28302).

\bibliographystyle{IEEEtran}
\bibliography{references}

\end{document}